\documentclass[a4paper,11pt]{article}
\pdfoutput=1 % 

\usepackage{jheppub} 
\usepackage{amsmath,amssymb,graphicx,float,slashed,xcolor,multicol}
\usepackage{tabularx}
\usepackage{url}
\usepackage{footmisc}
\usepackage{amsfonts}
\usepackage{cancel}
\usepackage{color}
\usepackage{multirow} 
\usepackage{pifont}
\usepackage{epstopdf}
\usepackage{comment}
\usepackage{booktabs} 
\usepackage{natbib}
\usepackage{array}
\usepackage{mathrsfs}
\usepackage[toc,page]{appendix}
\usepackage{mathtools}
\usepackage{romannum}
\usepackage{ulem}
\usepackage{bbold}
\usepackage{enumitem}
\usepackage{multirow}
\usepackage{cleveref}
\usepackage{caption,subcaption}
\usepackage{float}
\usepackage{multirow}
\usepackage{upgreek}
\usepackage[toc,page]{appendix}
\usepackage{romannum}
\allowdisplaybreaks
\usepackage{bbm}                % for \mathbbm{1} (unit matrix)
\usepackage{xspace}				% For spacing after command

\usepackage{tikz}
\usetikzlibrary{arrows,shapes}
\usetikzlibrary{trees}
\usetikzlibrary{matrix} 
\usetikzlibrary{positioning}				% For "above of=" commands
\usetikzlibrary{calc,through}				% For coordinates
\usetikzlibrary{decorations.pathreplacing}  % For curly braces
\usepackage{pgffor}							% For repeating patterns
\usetikzlibrary{decorations.pathmorphing}	% For Feynman Diagrams
\usetikzlibrary{decorations.markings}
\tikzset{
	vector/.style={decorate, decoration={snake}, draw},
	provector/.style={decorate, decoration={snake,amplitude=2.5pt}, draw},
	antivector/.style={decorate, decoration={snake,amplitude=-2.5pt}, draw},
	fermion/.style={draw=black, postaction={decorate},
		decoration={markings,mark=at position .55 with {\arrow[draw=black]{>}}}},
	fermionbar/.style={draw=black, postaction={decorate},
		decoration={markings,mark=at position .55 with {\arrow[draw=black]{<}}}},
	fermionnoarrow/.style={draw=black},
	gluon/.style={decorate, draw=black,
		decoration={coil,amplitude=4pt, segment length=5pt}},
	scalar/.style={dashed,draw=black, postaction={decorate},
		decoration={markings,mark=at position .55 with {\arrow[draw=black]{>}}}},
	scalarbar/.style={dashed,draw=black, postaction={decorate},
		decoration={markings,mark=at position .55 with {\arrow[draw=black]{<}}}},
	scalarnoarrow/.style={dashed,draw=black},
	electron/.style={draw=black, postaction={decorate},
		decoration={markings,mark=at position .55 with {\arrow[draw=black]{>}}}},
	bigvector/.style={decorate, decoration={snake,amplitude=4pt}, draw},
}

\tikzset{>=latex} % set arrow style

\usepackage{tikz-feynman}
\makeatletter
\tikzfeynmanset{compat=\tikzfeynman@version@major.\tikzfeynman@version@minor.\tikzfeynman@version@patch}
\makeatother

\tikzstyle{block} = [draw, rectangle, 
minimum height=3em, minimum width=6em]

\newcommand*{\rom}[1]{\expandafter\@slowromancap\romannumeral #1@}

\newcommand{\Q}{\mathcal {O}}
\newcommand{\C}{\mathcal {C}}

\def\lag{\mathscr{L}}

\def\us{\underset}

\def\beq{\begin{equation}}
\def\eeq{\end{equation}}
\def\beqa{\begin{eqnarray}}
\def\eeqa{\end{eqnarray}}
\title{Anomalous dimensions from gauge couplings in SMEFT with right-handed neutrinos}
\author[a]{Alakabha Datta,}
\author[b]{Jacky Kumar,}
\author[c]{Hongkai Liu,}
\author[d]{Danny Marfatia}
\affiliation[a]{Department of Physics and Astronomy, University of Mississippi, Oxford, MS 38677, USA}
\affiliation[b]{Physique des Particules, Universit´e de Montr´eal,
	C.P. 6128, succ. centre-ville, Montr´eal, QC, Canada H3C 3J7}
\affiliation[c]{Department of Physics and Astronomy, University of Pittsburgh, Pittsburgh, PA 15260, USA}
\affiliation[d]{Department of Physics and Astronomy, University of Hawaii at Manoa, Honolulu, HI 96822, USA}
\emailAdd{datta@phy.olemiss.edu}
\emailAdd{jacky.kumar@umontreal.ca}
\emailAdd{hol42@pitt.edu}
\emailAdd{dmarf8@hawaii.edu}

\preprint{
	\begin{flushright}
	\end{flushright}
}

\abstract{ 
	Standard Model Neutrino Effective Field Theory (SMNEFT) is an
	effective theory with Standard Model (SM) gauge-invariant operators constructed only from SM
	and right-handed neutrino fields.
	For the full set of dimension-six SMNEFT operators, we present the gauge coupling terms of the one-loop anomalous dimension matrix for renormalization 
	group evolution (RGE) of the Wilson coefficients between a new physics
	scale and the electroweak scale.
	We find that the SMNEFT operators can be divided into five subsets which are closed under RGE. Our results apply for both Dirac and Majorana neutrinos.
	We also discuss the operator mixing pattern numerically and comment on some interesting
	phenomenological implications. 
}

\begin{document}
	
	\titlepage
	\maketitle
	\newpage

	%%%%%%%%%%%%%%%%%%%%%%%%%%%
	
	\flushbottom
	
	%%%%%%%%%%%%%%%%%%%%%%%%%%%
	\section{Introduction}
	\label{sec:intro}
	The Standard Model (SM) of particle physics is an effective theory valid to some mass scale $\Lambda$. New physics at the scale $\Lambda$ may address
	important issues like the origin of the electroweak scale, $\mu_{EW}$. In the SM, 
	electroweak symmetry breaking arises from a complex
	fundamental Higgs scalar.  Between  $\mu_{EW}$ and  $\Lambda$, an effective field theory (EFT) framework can be used to describe new physics in a model independent way.
	In this approach, the leading 
	terms are given by the SM, and corrections from an underlying
	theory beyond the SM are described by higher dimension
	operators,
	\begin{eqnarray}
	{\cal L}  =  \sum_{i} {\cal{C}}_i {\cal O}_i\,. \ 
	\end{eqnarray}
	The operators ${\cal O}_i$ are
	$SU(3)_C \times SU(2)_L \times U(1)_Y$ invariant and are constructed only from SM fields. The Wilson coefficients (WCs)
	${\cal {C}}_i$,
	that determine the size of the contribution of operators ${\cal O}_i$,
	can be calculated by matching the effective theory with the underlying
	theory. 
	
	Analyses of higher dimension
	operators~\cite{Buchmuller:1985jz} have begun anew in the study of the SM as an effective field theory.
	Due to the phenomenological success of the SM gauge theory and the Higgs mechanism, the most studied EFT is the Standard Model Effective Field Theory (SMEFT)~\cite{Grzadkowski:2010es,Henning:2014wua, Brivio:2017vri}, which respects the SM gauge symmetry with only SM field content. The one-loop renormalization group evolution (RGE) of all dimension-six operators in SMEFT  have been calculated in Refs.~\cite{Jenkins:2013zja, Jenkins:2013wua, Alonso:2013hga}. 
	
	In the SMEFT framework, new physics is considered to be heavy  with  $ \Lambda \gg \mu_{EW}$.  However, many experiments point to new physics with a mass scale well below the electroweak scale,
	and many  experiments to search for new light states are planned.
	Since these states do not appear in SMEFT, its Lagrangian must be supplemented by interactions between these new states and the SM fields. 
	Possible new states are right-handed neutrinos that are sterile under SM gauge interactions. The masses of the sterile neutrinos can vary over a large range and can be heavy or light compared to the electroweak scale. Light sterile neutrinos have been invoked to explain many phenomena; see Ref.~\cite{Abazajian:2012ys} for a review.
	
	In this paper, we consider the sterile neutrinos to be light so that they appear as explicit degree of freedoms in the EFT framework.  %Light right-handed neutrinos have been considered as  warm dark matter candidates~\cite{Dodelson:1993je}  and a portal to  new physics~\cite{Escudero:2016ksa}. 
	% They have been proposed to explain the MiniBooNe excess electron like events \cite{Bertuzzo:2018itn, Ballett:2018ynz, Datta:2020auq}
	% and explain the   $ R(D^{(*)}) $ measurements 
	% in semileptonic $B \to  D^{(*)} \tau \bar{\nu}_\tau$  decays \cite{He:2012zp, Asadi:2018wea, Greljo:2018ogz, Babu:2018vrl, Mandal:2020htr}.
	%  Light sterile neutrino of the Dirac nature, like all other SM fermions, is the simplest explanation of neutrino masses. Dirac neutrino with no BSM interactions except for Yukawa interactions is very challenging for phenomenological study and so other interactions of the sterile neutrino with the SM fields should be considered.
	% Considering the new physics in need in the neutrino sector, it is crucial to study the potential new interactions in the neutrino sector.
	%Massive neutrinos is one of the main motivation for new physics in the neutrino sector. Given the fact that there is no direct BSM signals produced at the high-energy experiment, like %LHC, it is natural to include all the possible interactions in a consistent way. 
We use the Standard Model Neutrino Effective Field Theory (SMNEFT)  which augments SMEFT with right-handed (RH) neutrinos $n$~\cite{delAguila:2008ir, Aparici:2009fh, Bhattacharya:2015vja, Liao:2016qyd,Bischer:2019ttk}. The RGE of some SMNEFT operators have been calculated. The mixing between the bosonic operators has been calculated in Refs.~\cite{Bell:2005kz, Chala:2020pbn}, and the one-loop RGE of a subset of four-fermion operators are given in Ref.~\cite{Han:2020pff}. In this work, we present the gauge terms of the one-loop RGE of all dimension-six operators in SMNEFT. 
	
	The paper is organized as follows. In section~2, we define SMNEFT and establish our notation. Our diagrammatic approach to calculate the one-loop anomalous dimension
	matrix (ADM) is described in section~3. In section~4, we present the ADM. In section~5, we study operator mixing using the leading-log approximation. 
	We discuss some phenomenological implications in section~6, and summarize in section~7. Details of our calculations are provided in an appendix.\\
	
	\section{SMNEFT}
	In this section, we present SMNEFT. Neutrinos may be Dirac or Majorana. In the case of Dirac neutrinos, $\nu_R \equiv n$, with $n$ and the left-handed neutrino $\nu_L$ in the same spinor $\nu_D = (\nu_L, n)^T$, and of the same mass. In the Majorana case, $n$ and $\nu_L$ are components of two different spinors, $\nu_M=(\nu_L,\nu_L^c)^T$, $n_M = (n^c,n)^T$, and can have different masses. Our results are valid for both cases because we focus on the gauge sector. Without specifying any possible Majorana and Dirac mass terms, the dimension-six $B$ and $L$ conserving SMNEFT Lagrangian is
	\beq
	\lag^{(6)}_{\text{SMNEFT}} \supset \lag_{\rm{SM}} + i \bar{n}\slashed{\partial}n
	+ \sum_{i} \mathcal{C}_i\mathcal{O}_i\,,
	\eeq
	where $\mathcal{C}_i$ are the WCs with the scale of new physics absorbed in them, and the 
	SM Lagrangian is given by
	\begin{eqnarray}
	\lag_{\rm SM}= &=& 
	-\frac{1}{4} G_{\mu \nu}^A G^{A \mu \nu}
	-\frac{1}{4} W_{\mu \nu}^I W^{I \mu \nu}
	-\frac{1}{4} B_{\mu \nu} B^{\mu \nu} \notag \\
	&+& (D_\mu \phi)^\dagger (D^\mu\phi)
	+ m^2 \phi^\dagger \phi - \frac{\lambda}{2} (\phi^\dagger \phi)^2 \notag \\
	&+& i(\bar \ell \slashed{D}\ell + \bar e \slashed{D}e
	+ \bar q \slashed{D} q + \bar u \slashed{D} u 
	+ \bar d \slashed{D} d) \notag \\
	&-&(\bar \ell Y_e e \phi + \bar q Y_u u \tilde {\phi}
	+ \bar q Y_d d \phi + \rm{h.c.})\,.
	\end{eqnarray}
	%The Yukawa term for the RH neutrino is
	%\beq
	%\lag^n_{\rm{Yukawa}} = - \bar \ell  Y_n n \tilde \varphi + %\rm{h.c.},
	%\eeq
	%where $\tilde{\varphi} \equiv i\tau_2 \varphi^*$.
	Here, $\tilde{\phi}^j = \epsilon_{jk} (\phi^k)^*$, and the Higgs vacuum expectation value is 
	$\langle\phi\rangle = v/\sqrt{2}$ with $v=246$~GeV.  
	The covariant derivative and field strength tensors are defined by
	\beq
	D_\mu= \partial_\mu + i g_1 y B_\mu + ig_2 \frac{\tau^I}{2} W^I_\mu + ig_3 \frac{T^a}{2} G^a_\mu\,,
	\eeq
	\beqa
	B_{\mu\nu} &=& \partial_\mu B_\nu - \partial_\nu B_\mu\,,\\
	W^I_{\mu\nu}&=& \partial_\mu W_\nu^I - \partial_\nu W_\mu^I - g_2 \epsilon^{IJK}W_\mu^JW_\nu^K\,,\\
	G^a_{\mu\nu}&=& \partial_\mu G_\nu^a - \partial_\nu G_\mu^a - g_3 f^{abc}G_\mu^bG_\nu^c\,,
	\eeqa
	where $g_1$, $g_2$, and $g_3$ are the gauge couplings of $U(1)_Y$, $SU(2)_L$, and $SU(3)_C$, 
	respectively, and $y$ is the hypercharge. $\epsilon^{IJK}$ and $f^{abc}$ are the $SU(2)_L$ and $SU(3)_C$ structure constants, respectively. 
	
	The 16 baryon and lepton number conserving 
	($\Delta B = \Delta L $ =0 ) operators 
	involving the field $n$ in SMNEFT  are 
	shown in Table~\ref{Tb:SMNEFT}~\cite{Liao:2016qyd} in the WCxf convention~\cite{Aebischer:2017ugx}. 
	
	\begin{table}
		\centering
		\renewcommand{\arraystretch}{1.5}
		\begin{tabular}{|c|c|c|c|c|c|}
			\hline
			\multicolumn{2}{|c|}{$(\bar RR)(\bar RR)$} &
			\multicolumn{2}{|c|}{$(\bar LL)(\bar RR)$} &
			\multicolumn{2}{|c|}{$(\bar LR)(\bar RL)$ and $(\bar LR)(\bar LR)$ }\\
			\hline
			$\Q_{nd}$  & $(\bar n_p \gamma_\mu n_r)(\bar d_s \gamma^\mu d_t)$ &
			$\Q_{qn}$  & $(\bar q_p \gamma_\mu q_r)(\bar n_s \gamma^\mu n_t)$ &
			$\Q_{\ell n \ell e}$  & $(\bar \ell^j_p  n_r)\epsilon_{jk}(\bar \ell^k_s e_t)$   \\
			$\Q_{nu}$  & $(\bar n_p \gamma_\mu  n_r)(\bar u_s \gamma^\mu  u_t)$ &
			$\Q_{\ell n}$ & $(\bar \ell_p \gamma_\mu \ell_r)(\bar n_s\gamma^\mu n_t)$ &
			$\Q_{\ell n q d}^{(1)}$ &  $(\bar \ell^j_p  n_r)\epsilon_{jk}(\bar q^k_s  d_t)$  \\
			$\Q_{ne}$  & $(\bar n_p \gamma_\mu n_r)(\bar e_s \gamma^\mu e_t)$ &
			& &
			$\Q_{\ell n q d}^{(3)}$  & $(\bar \ell^j_p  \sigma_{\mu\nu} n_r)\epsilon_{jk}(\bar q^k_s \sigma^{\mu\nu} d_t)$     \\
			$\Q_{nn}$  & $(\bar n_p \gamma_\mu n_r)(\bar n_s \gamma^\mu  n_t)$ &
			&  &
			$\Q_{\ell n u q}$  & $(\bar \ell^j_p  n_r)(\bar u_s q^j_t)$  \\
			$\Q_{nedu}$      & $(\bar n_p \gamma_\mu e_r)(\bar d_s \gamma^\mu  u_t)$   & & &
			&   \\
			\hline
			\multicolumn{2}{|c|}{$\psi^2\phi^3$} &
			\multicolumn{2}{|c|}{$\psi^2\phi^2 D$} &
			\multicolumn{2}{|c|}{$\psi^2 X \phi$}\\
			\hline
			$\Q_{n\phi}$  & $(\phi^\dagger \phi) (\bar l_p  n_r \tilde \phi)$ & $\Q_{\phi n}$   &
			$i(\phi^\dagger \overset{\leftrightarrow}{D}_\mu \phi) (\bar n_p \gamma^\mu n_r) $& $\Q_{nW}$ & $(\bar \ell_p  \sigma^{\mu\nu} n_r)\tau^I \tilde \phi W_{\mu\nu}^I$ \\
			&  & $\Q_{\phi ne}$   &  $i(\tilde \phi^\dagger D_\mu \phi) (\bar n_p \gamma^\mu e_r) $  & $\Q_{nB}$  & $(\bar \ell_p  \sigma^{\mu\nu} n_r)\tilde \phi B_{\mu\nu}$ \\
			&  &  &  &  &  \\
			\hline
		\end{tabular}
		\bigskip
		\captionsetup{width=0.9\textwidth}
		\caption{\small \label{tab:SMNEFTops} The 16 SMNEFT operators involving RH neutrinos $n$ in the 
			Warsaw basis convention which conserve baryon and lepton number $(\Delta B=\Delta L=0)$. 
			The flavor indices $`prst`$ on the operators are suppressed for simplicity. The fundamental $SU(2)_L$ 
			indices are denoted by $i,j$, and $I$ is the adjoint index. }
		\label{Tb:SMNEFT}
	\end{table}

	%%%%%%%%%%%%%%%%%%%%%%%%%%%%%%%%%%%%
	\section{Formalism}
	\label{sec:derivations}
	The anomalous dimensions of an operator are given by the infinite pieces, i.e., the coefficients of the $1/\varepsilon$ terms of the diagrams.  
	In this section, we define our procedure to calculate the ADM, and relegate the details of our calculations to Appendix~\ref{app:derivations}. 
	% the renormalization constants for th we compute the ADMs for the full set of SMNEFT operators as 
	%shown in Table \ref{Tb:SMNEFT}. 
	To compute the ADM we use the master formulae presented in Ref.~\cite{Buras:1998raa}. 
	We compute one-loop contributions to the ADM due to SM gauge couplings.
	The four-fermion operators ($\psi^4$) in Table~\ref{Tb:SMNEFT} can be divided
	into four categories:
	$(\bar R R)(\bar R R)$, $(\bar LL)(\bar R R)$, $(\bar L R)(\bar R L)$, and 
	$(\bar L R)(\bar L R)$ on the basis of the chiralities of the fields. 
	The remaining operators are of the form $\psi^2 \phi^3$, $\psi^2 \phi^2 D$ 
	and $\psi^2 X \phi$. We focus on the 
	$\psi^4$-$\psi^4$ and $\psi^4$ - $\psi^2 \phi^2 D$ operator mixing since the mixing between 
	$\psi^2\phi^3$, $\psi^2 \phi^2 D$ and $\psi^2 X \phi$ has been 
	computed in the Ref.~\cite{Chala:2020pbn} using the background field method.
	We have checked that the resulting $5\times 5 $ matrix is consistent 
	with the result for the corresponding SMEFT operators~\cite{Alonso:2013hga} which have a similar 
	ADM structure. 
	
	For the $\psi^4$ operators the bare and renormalized operators are related by
	\\
	\begin{equation}
	\label{eq:op:ren}
	\langle {\vec{\mathcal{O}}} \rangle^{(0)} = 
	{{Z_{\psi_1}^{-\frac{1}{2}}} {Z_{\psi_2}^{-\frac{1}{2}}} {Z_{\psi_3}^{-\frac{1}{2}}} 
		{Z_{\psi_4}^{-\frac{1}{2}}}} ~{\hat Z} \langle \vec{\mathcal{O}} \rangle  = Z \langle \vec{\mathcal{O}} \rangle\,,
	\end{equation}
	where the superscript $(0)$ labels the bare  matrix elements. Here, $\hat{Z}$ and $Z_{\psi}$ 
	are the renormalization constants for the operator $\vec{\mathcal{O}}$ and the  
	fields $\psi$, respectively. In the $\overline{\rm MS}$ scheme at one-loop level, 
	the renormalization constants take the form,
	\\
	\begin{eqnarray}
	\label{eq:constants:fields}
	(Z_m)_{\psi} &=& 1 + \frac{\alpha_m}{4\pi}  \frac{1}{\varepsilon} a^m_\psi\,,  \\
	\label{eq:constants:full}
	(Z_m)_{ij} &=& \delta_{ij} + \frac{\alpha_m}{4\pi} \frac{1}{\varepsilon} b^m_{ij}\,, \\
	\label{eq:constants:ops}
	(\hat{Z}_m)_{ij} &=& \delta_{ij}  +  \frac{\alpha_{m}}{4\pi} \frac{1}{\varepsilon} c^m_{ij}\,,
	\end{eqnarray}
	with $\psi=  \{q,u,d, \ell, e\}$ and $Z_n=1$. The coupling constants are defined by
	$\alpha_m={g_m^2}/{4\pi}$ with $m=1$, 2, 3 
	for $U(1)_Y$, $SU(2)_L$ and $SU(3)_C$, respectively. 
	The coefficients of the UV divergent parts of the diagrams ($\alpha_{m}/(4\pi\varepsilon)$),  
	$a^m_\psi$, $b^m_{ij}$ and $c^m_{ij}$, are independent of the gauge couplings. 
	Note that $c^m_{ij}$ can be related to $a^m_\psi$ and $b^m_{ij}$ via Eq.~\eqref{eq:op:ren}. 
	The anomalous dimension matrices are defined by the RG equations,
	\\
	%\beq
	%\dot {\mathcal C}_i (\mu) = 16 \pi^2 \mu \frac{d}{d\mu}{\cal C}_{i}(\mu) = \gamma_{ji} {\cal C}_{j}(\mu),
	%\eeq
	\beq
	\dot {\mathcal C}_i (\mu) = 16 \pi^2 \mu \frac{d}{d\mu}{\cal C}_{i}(\mu) = {(\gamma_\C)}_{ij} {\cal C}_{j}(\mu),
	\eeq
	where $\gamma_\C = \gamma^T$ with $\gamma$ given by the matrix ${\hat{Z}}$ as
	\begin{equation}
	\frac{\gamma }{16\pi^2}=  {\hat Z}^{-1} \frac{d {\hat Z}} {d ~{\rm \ln}~ \mu}\,,
	\end{equation}
	and which can be directly expressed in terms of $a^m_\psi$ and $b^m_{ij}$:
	\begin{equation}
	\label{eq:adms-master}
	\gamma_{ij} = -2 g_m^2 \left( \sum_{\psi = \psi_1, ..\psi_4} \frac{1}{2}a^m_\psi \delta_{ij} + b^{m}_{ij}  \right ) 
	\,.
	\end{equation}
	Here, the sum is over external fields $\psi_1$ to $\psi_4$ in a given operator, and summation over $m$ is implicit.
	Therefore, in order to compute the ADM for  a set of operators, we need to calculate the coefficients
	$a^m_\psi$ and $b^m_{ij}$ from the field strength renormalization and operator renormalization, 
	respectively.

\begin{figure}[t]
		\centering
		\begin{subfigure}{.25\textwidth}
			\centering
			\includegraphics[width=\textwidth]{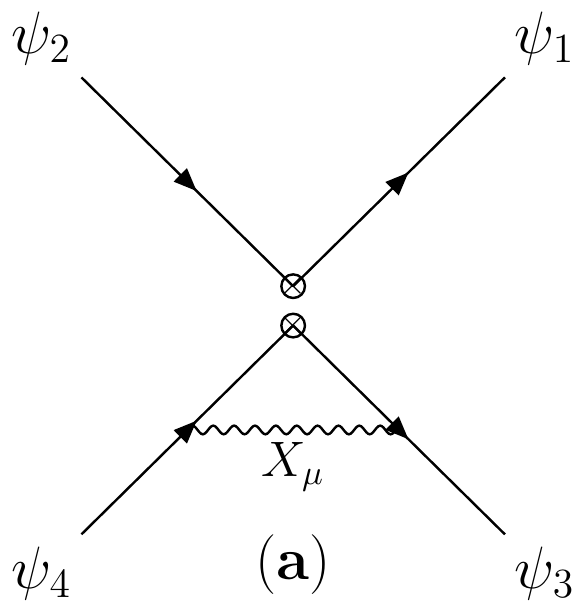}
		\end{subfigure}
		\begin{subfigure}{.25\textwidth}
			\centering
			\includegraphics[width=\textwidth]{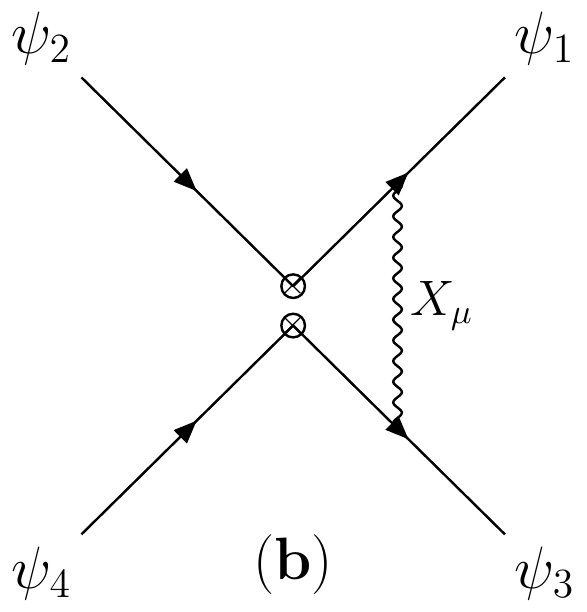}
		\end{subfigure}
		\begin{subfigure}{.25\textwidth}
			\centering
			\includegraphics[width=\textwidth]{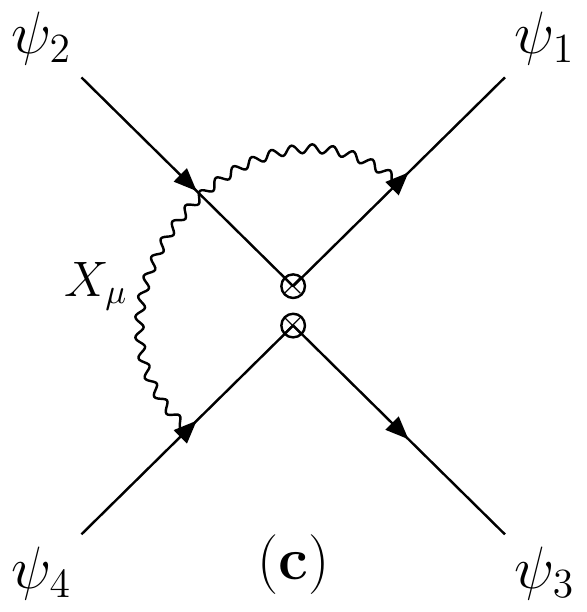}
		\end{subfigure}
		\caption{Current-current topologies with four-fermion insertions. Here $X_\mu$ represents the gauge bosons $B_\mu$, $W_\mu$ and $G_\mu$. The fermion fields $q,u,d, \ell, e$ and $n$ 
			are represented by $\psi_I$.
		}
		\label{fig:current-current}
	\end{figure}

\begin{figure}[t]
		\centering
		\begin{subfigure}{.2\textwidth}
			\centering
			\includegraphics[width=\textwidth]{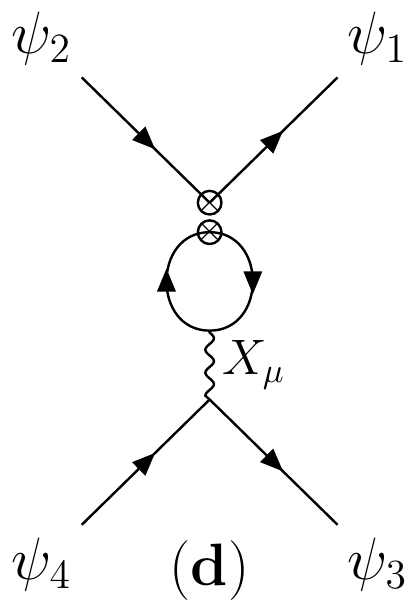}
		\end{subfigure}
		\begin{subfigure}{.2\textwidth}
			\centering
			\includegraphics[width=\textwidth]{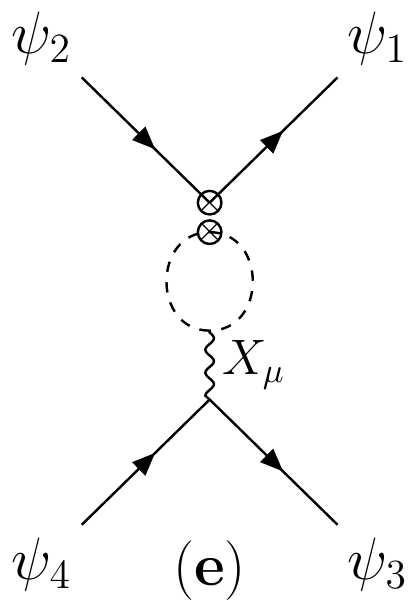}
		\end{subfigure}
		\caption{Penguin topologies with four-fermion (d) and boson (e) insertions. 
		}
		\label{fig:penguin}
	\end{figure}

	For the mixing between $\psi^4$-$\psi^4$ and $\psi^4$-$\psi^2 \phi^2 D$, 
	the current-current (Fig.~\ref{fig:current-current}) and penguin (Fig.~\ref{fig:penguin}) topologies mediated by the
	gauge bosons $X_\mu=B_\mu,W_\mu, G_\mu$, or the scalar, have to be calculated. These diagrams can be 
	computed by easily generalizing the master formulae of Ref.~\cite{Buras:1998raa} to SMNEFT; see Eqs.~(\ref{eq:div:cc})-(\ref{eq:div:penguin}).
	
	In Appendix~\ref{app:derivations}, we present explicit calculations of the ADMs for
	$\mathcal{O}_{nd}-\mathcal{O}_{nd}$, 
	$\mathcal{O}_{\phi n}-\mathcal{O}_{nd}$,
	$\mathcal{O}_{nedu}-\mathcal{O}_{nedu}$ and 
	$\mathcal{O}_{\ell n\ell e}-\mathcal{O}_{\ell n \ell e}$
	operator mixing. The same method is applicable to the other operators. It is worth noting that for most of the cases the structure of the 
	ADMs of the SMNEFT operators are similar to those of 
	SMEFT operators~\cite{Alonso:2013hga}. Therefore, our SMNEFT 
	results also serve as an important 
	cross-check for the corresponding gauge terms appearing in the SMEFT ADMs.

	\section{Anomalous dimensions}
	\label{sec:results}
	We now present terms for the one-loop ADM that depend on the gauge couplings
	$\alpha_1$, $ \alpha_2$ and $\alpha_3$ for all 16 SMNEFT operators. 
	The general formula for the ADM is given by Eq.~\eqref{eq:adms-master} and details 
	of the calculations of the Feynman diagrams to extract $a^m_{\psi}$ and $b^{m}_{ij}$ can be found in Appendix~\ref{app:derivations}. 
	The ADM for bosonic SMNEFT operators is given in Ref.~\cite{Chala:2020pbn}. The ADM of most SMNEFT operators can be obtained from the ADM of the SMEFT operators~\cite{Alonso:2013hga} with a similar structure. For example, the ADM for the SMNEFT operators $\Q_{\ell nuq}$, $\Q^{(1)}_{\ell nqd}$ and $\Q^{(3)}_{\ell nqd}$, can be obtained by replacing $e$ with $n$, and switching $u$ and $d$ in the SMEFT operators $\Q_{\ell edq}$, $\Q^{(1)}_{\ell equ}$ and $\Q^{(3)}_{\ell equ}$. We use this procedure as a cross-check when available. No such comparison is possible for $\Q_{nedu}$, which has a structure not present in SMEFT.
	\begin{boldmath}
		\subsection{$\psi^4$}
	\end{boldmath}
	
	The ADM for four-fermion operators are provided below. 
	
	\begin{boldmath}
		\subsubsection{$(\bar RR)(\bar RR)$}
	\end{boldmath}
	\beqa
	\dot \C_{\underset{prst}{nd}} &=& ( \frac{4}{3} N_c y_d^2 \C_{\underset{prww}{nd}} +\frac{4}{3} N_c y_dy_u \C_{\underset{prww}{nu}}+ \frac{4}{3} y_dy_e \C_{\underset{prww}{ne}}+ \frac{8}{3} N_c y_d y_q \C_{\underset{wwpr}{q n}}
	+\frac{8}{3} y_d y_{\ell} \C_{\underset{wwpr}{\ell n}} \notag \\
	&\quad&+\frac{4}{3}y_dy_h \C_{\underset{pr}{\phi n}}) {g_1^2}\delta_{st}\,, \\
	\dot \C_{\underset{prst}{nu}} &=& (\frac{4}{3} N_c y_uy_d \C_{\underset{prww}{nd}}+ \frac{4}{3} N_c y_u^2 \C_{\underset{prww}{nu}}+ \frac{4}{3} y_uy_e \C_{\underset{prww}{ne}} + \frac{8}{3} N_c y_u y_q \C_{\underset{wwpr}{q n}}
	+\frac{8}{3} y_u y_{\ell} \C_{\underset{wwpr}{\ell n}} \notag \\
	&\quad&+\frac{4}{3}y_uy_h \C_{\underset{pr}{\phi n}}) {g_1^2} \delta_{st}\,,\\
	\dot \C_{\underset{prst}{ne}} &=& (\frac{4}{3} N_c y_ey_d \C_{\underset{prww}{nd}} + \frac{4}{3} N_c y_ey_u \C_{\underset{prww}{nu}}+ \frac{4}{3} y_e^2 \C_{\underset{prww}{ne}} + \frac{8}{3} N_c y_e y_q \C_{\underset{wwpr}{q n}}
	+\frac{8}{3} y_e y_{\ell} \C_{\underset{wwpr}{\ell n}}\notag \\
	&\quad&+\frac{4}{3}y_ey_h \C_{\underset{pr}{\phi n}}) {g_1^2}\delta_{st}\,,\\ 
	\dot \C_{\underset{prst}{nedu}} &=& ( (y_d-y_u)^2 + y_e (y_e+8 y_u - 2 y_d)) {g_1^2} \C_{\underset{prst}{nedu}}\,,\\
	\dot \C_{\underset{prst}{nn}} &=& 0\,,
	\eeqa
	
	\begin{boldmath}
		\subsubsection{$(\bar LL)(\bar RR)$}
	\end{boldmath}
	
	\beqa
	\dot \C_{\underset{prst}{qn}} &=&(\frac{4}{3} N_c y_qy_d \C_{\underset{stww}{nd}} + \frac{4}{3} N_c y_qy_u \C_{\underset{stww}{nu}}+ \frac{4}{3} y_qy_e \C_{\underset{stww}{ne}} + \frac{8}{3} N_c y_q^2 \C_{\underset{wwst}{qn}}
	+\frac{8}{3} y_q y_{\ell} \C_{\underset{wwst}{\ell n}} \notag \\
	&\quad&+\frac{4}{3}y_qy_h \C_{\underset{st}{\phi n}}) {g_1^2}\delta_{pr}\,,\\
	\dot \C_{\underset{prst}{\ell n}} &=&( \frac{4}{3} N_c y_{\ell}y_d \C_{\underset{stww}{nd}} + \frac{4}{3} N_c y_{\ell}y_u \C_{\underset{stww}{nu}}+ \frac{4}{3} y_{\ell}y_e \C_{\underset{stww}{ne}}+ \frac{8}{3} N_c y_{\ell}y_q \C_{\underset{wwst}{qn}}
	+\frac{8}{3} y_{\ell}^2 \C_{\underset{wwst}{\ell n}}  \notag \\
	&\quad&+\frac{4}{3}y_{\ell}y_h \C_{\underset{st}{\phi n}}) {g_1^2}\delta_{pr}\,.
	\eeqa

	\begin{boldmath}
		\subsubsection{$(\bar L R)(\bar R L)$ and $(\bar L R)(\bar L R)$}
	\end{boldmath}
	
	\beq
	\dot \C_{\underset{prst}{\ell n\ell e}} = ((y_e^2-8y_ey_\ell+6y_\ell^2) {g_1^2} -\frac{3}{2}  {g_2^2})\C_{\underset{prst}{\ell n\ell e}} - (4 y_\ell(y_e+y_\ell) {g_1^2} -3  {g_2^2} )\C_{\underset{srpt}{\ell n\ell e}}\,,
	\eeq
	\beqa
	\dot \C^{(1)}_{\underset{prst}{\ell nqd}} &=& ((y_d^2-2y_d(y_{\ell}+4y_q)+ (y_{\ell} + y_q)^2) {g_1^2} -8 {g_3^2} )C^{(1)}_{\underset{prst}{\ell nqd}}\,, \notag \\
	&\quad&+ (-24 y_{\ell} (y_d+y_q) {g_1^2}  + 18 {g_2^2}  )C^{(3)}_{\underset{prst}{\ell nqd}}\,,\\
	\dot \C^{(3)}_{\underset{prst}{\ell nqd}} &=&(-\frac{1}{2}y_{\ell} (y_d+y_q) {g_1^2}  + \frac{3}{8} {g_2^2}   )C^{(1)}_{\underset{prst}{\ell nqd}} \notag\\
	&\quad&+ ((y_d^2-6y_dy_{\ell}+y_{\ell}^2+6y_{\ell}y_q+y_q^2) {g_1^2} -3 {g_2^2}+\frac{8}{3} {g_3}^2 )C^{(3)}_{\underset{prst}{\ell nqd}}\,,\\
	\dot \C_{\underset{prst}{\ell nuq}} &=& ( ((y_{\ell}+y_{u})^2+y_q(y_q-2y_\ell-8y_u)) {g_1^2}-8 {g_3^2} )\C_{\underset{prst}{\ell nuq}}\,.
	\eeqa

	\begin{boldmath}
		\subsection{$\psi^2 \phi^3$}
	\end{boldmath}
	
	\beq
	\begin{split}
		\dot \C_{\underset{pr}{n\phi}} =&-(9y_{\ell}^2 {g_1^2}+\frac{27}{4} {g_2^2})\C_{\underset{pr}{n\phi}} - 6(4y_h^2y_{\ell} {g_1^3}-y_h {g_1g_2^2})\C_{\underset{pr}{nB}}\\
		&+3(4y_hy_{\ell} {g_1^2g_2}-3 {g_2^3})\C_{\underset{pr}{nW}}\,.
	\end{split}
	\eeq
	
	\begin{boldmath}
		\subsection{$\psi^2 \phi^2 D$}
	\end{boldmath}
	\beqa
	\dot \C_{\underset{pr}{\phi n}} &=&(\frac{4}{3} y_h^2 \C_{\underset{pr}{\phi n}}+ \frac{4}{3} N_c y_dy_h \C_{\underset{prww}{nd}} +\frac{4}{3} N_c y_uy_h \C_{\underset{prww}{nu}}+ \frac{4}{3} y_ey_h \C_{\underset{prww}{ne}}+\frac{8}{3}N_c y_qy_h \C_{\underset{wwpr}{q n}} \notag \label{eq:adm_phin}\\
	&\quad&+ \frac{8}{3} y_{\ell} y_h \C_{\underset{wwpr}{\ell n}}) {g_1^2}\,,\\
	\dot \C_{\underset{pr}{\phi ne}} &=&(-3 y_e^2 \C_{\underset{pr}{\phi ne}}) {g_1^2} \label{eq:adm_phine}\,.
	\eeqa
	
	\begin{boldmath}
		\subsection{$\psi^2 X \phi$}
	\end{boldmath}
	
	\beqa
	\dot \C_{\underset{pr}{nW}} &=& ((3 \C_{F,2} - b_{0,2}) {g_2^2} - 3y_{\ell}^2 {g_1^2})\C_{\underset{pr}{nW}} + 3y_{\ell} {g_1g_2}\C_{\underset{pr}{nB}}\,,\\ 
	\dot \C_{\underset{pr}{nB}} &=& (-3 \C_{F,2} {g_2^2} + (3y_{\ell}^2-b_{0,1}){g_1^2})\C_{\underset{pr}{nB}} + 12 \C_{F,2}y_{\ell} {g_1g_2}C_{\underset{pr}{nW}}\,, 
	\eeqa
	where the quadratic Casimir $C_{F,2} = \frac{3}{4}$. $b_{0,1} = -\frac{41}{6}$ and $b_{0,2} = \frac{19}{6}$  are the first coefficients in the $g_1$ and $g_2$ $\beta- $functions, respectively. 
	\\
	\\
	
	\section{Operator mixing}
	\label{sec:discussion}
	We study operator mixing by solving the RG equations presented above in the leading-log 
	approximation. The solution to these equations for running between scales $\Lambda$ and $\mu$ 
	is 
	\begin{equation}
	\mathcal{C}_{i}(\mu) = \left ( \delta_{ij} + \frac{{(\gamma_\C)}_{ij}}{16 \pi^2}
	\ln  \frac{\mu}{\Lambda} \right ) \mathcal{C}_{j}(\Lambda)\,.
	\label{eq:ll}
	\end{equation}
	
	Depending upon the mixing structure the operators are divided into five subsets 
	forming $6\times 6$, $3\times 3$, $3\times 3$, $2\times 2$, and $2\times 2$ ADMs. %The flavor structures are taken care of in the Section~\ref{sec:pheno}. 
	Defining $\delta \C_i(\mu) = \C_i(\mu) - \C_i(\Lambda)$, the leading-log solution for the first group reads
	\beq
	\begin{pmatrix}
		\delta \C_{\us{prst}{nd}}\\
		\delta \C_{\us{prst}{nu}}\\
		\delta \C_{\us{prst}{ne}}\\
		\delta \C_{\us{stpr}{qn}}\\
		\delta \C_{\us{stpr}{\ell n}}\\
		\delta \C_{\us{pr}{\phi n}}
	\end{pmatrix}_{(\mu)} = \frac{\alpha_1}{4\pi} \ln  \frac{\mu}{\Lambda}  \begin{pmatrix}
		\frac{4}{9}\delta_{st} & -\frac{8}{9}\delta_{st} & \frac{4}{9} \delta_{st} & -\frac{4}{9} \delta_{st} & \frac{4}{9} \delta_{st} & -\frac{2}{9} \delta_{st} \\
		-\frac{8}{9}\delta_{st} & \frac{16}{9}\delta_{st} & -\frac{8}{9} \delta_{st} & \frac{8}{9} \delta_{st} & -\frac{8}{9} \delta_{st} & \frac{4}{9} \delta_{st}\\
		\frac{4}{3}\delta_{st} & -\frac{8}{3}\delta_{st} & \frac{4}{3}\delta_{st} & -\frac{4}{3}\delta_{st} & \frac{4}{3}\delta_{st} & -\frac{2}{3}\delta_{st}\\
		-\frac{2}{9}\delta_{st} & \frac{4}{9}\delta_{st} & -\frac{2}{9}\delta_{st} & \frac{2}{9}\delta_{st} & -\frac{2}{9}\delta_{st} & \frac{1}{9}\delta_{st}\\
		\frac{2}{3}\delta_{st} & -\frac{4}{3}\delta_{st} & 	\frac{2}{3}\delta_{st} & -\frac{2}{3}\delta_{st} & \frac{2}{3}\delta_{st} & -\frac{1}{3}\delta_{st}\\
		-\frac{2}{3} & \frac{4}{3} & -2 & \frac{2}{3} & -\frac{2}{3} & \frac{1}{3}\\
	\end{pmatrix}
	\begin{pmatrix}
		\C_{\us{prww}{nd}}\\
		\C_{\us{prww}{nu}}\\
		\C_{\us{prww}{ne}}\\
		\C_{\us{wwpr}{qn}}\\
		\C_{\us{wwpr}{\ell n}}\\
		\C_{\us{pr}{\phi n}}
	\end{pmatrix}_{(\Lambda)}\,.
	\label{eq:adm_c1}
	\eeq
	Summation over the repeated $w$ index is implicit. Next, we have the $3 \times 3$ structure,
	\beq
	\begin{pmatrix}
		\delta	\C_{\us{pr}{n\phi}}\\
		\delta	\C_{\us{pr}{nW}}\\
		\delta	\C_{\us{pr}{nB}}
	\end{pmatrix}_{(\mu)} = \frac{1}{16\pi^2} \ln  \frac{\mu}{\Lambda}  \begin{pmatrix}
		-\frac{9}{4}(g_1^2+3g_2^2) & -3 g_2(g_1^2+3g_2^2) & 3 g_1(g_1^2 + g_2^2) \\
		0 & -\frac{1}{12}(9g_1^2+11g_2^2)& -\frac{3g_1g_2}{2}  \\
		0 & -\frac{9g_1g_2}{2} & \frac{9}{4} (\frac{91}{27}g_1^2-g_2^2)
	\end{pmatrix}
	\begin{pmatrix}
		\C_{\us{pr}{n\phi}}\\
		\C_{\us{pr}{nW}}\\
		\C_{\us{pr}{nB}}
	\end{pmatrix}_{(\Lambda)}.
	\label{eq:adm_c2}
	\eeq
	The operators $\C^{(1)}_{\ell nqd}$ and $\C^{(3)}_{\ell nqd}$ mix according to
	\beq
	\begin{pmatrix}
		\delta	\C^{(1)}_{\us{prst}{\ell nqd}}\\
		\delta	\C^{(3)}_{\us{prst}{\ell nqd}}
	\end{pmatrix}_{(\mu)} =\ln \frac{\mu}{\Lambda} [\frac{\alpha_1}{4\pi}  \begin{pmatrix}
		\frac{1}{3} & -2 &  \\
		-\frac{1}{24} & -\frac{10}{9}
	\end{pmatrix} + \frac{\alpha_2}{4\pi}\begin{pmatrix}
		0 & 18 &  \\
		\frac{3}{8} & -3
	\end{pmatrix} +\frac{\alpha_3}{4\pi}\begin{pmatrix}
		-8 & 0 &  \\
		0 & \frac{8}{3}
	\end{pmatrix} ]
	\begin{pmatrix}
		\C^{(1)}_{\us{prst}{\ell nqd}}\\
		\C^{(3)}_{\us{prst}{\ell nqd}}
	\end{pmatrix}_{(\Lambda)}.
	\label{eq:adm_c3}
	\eeq
	The operator $\C_{\ell n\ell e}$ mix with different flavors:
	\beq
	\begin{pmatrix}
		\delta \C_{\us{prst}{\ell n\ell e}}\\
		\delta \C_{\us{srpt}{\ell n\ell e}}\\
	\end{pmatrix}_{(\mu)} =\frac{\alpha_1+\alpha_2}{4\pi} \ln \frac{\mu}{\Lambda} \begin{pmatrix}
		-\frac{3}{2} &  -3  \\
		-3  &-\frac{3}{2}   \\
	\end{pmatrix} 
	\begin{pmatrix}
		\C_{\us{prst}{\ell n\ell e}}\\
		\C_{\us{srpt}{\ell n\ell e}}\\
	\end{pmatrix}_{(\Lambda)}.
	\label{eq:adm_c4}
	\eeq
	The remaining operators do not mix:
	\beq
	\begin{pmatrix}
		\delta \C_{\us{prst}{nedu}}\\
		\delta \C_{\us{prst}{\ell nuq}}\\
		\delta \C_{\us{pr}{\phi ne}}
	\end{pmatrix}_{(\mu)} = \ln \frac{\mu}{\Lambda} [\frac{\alpha_1}{4\pi}\begin{pmatrix}
		-4 &  & &  \\
		&  &-\frac{2}{3} &  \\
		&  & & -3 
	\end{pmatrix}  +\frac{\alpha_3}{4\pi}\begin{pmatrix}
		0 &  & &  \\
		&  &-8 &  \\
		&  & & 0
	\end{pmatrix} ]
	\begin{pmatrix}
		\C_{\us{prst}{nedu}}\\
		\C_{\us{prst}{\ell nuq}}\\
		\C_{\us{pr}{\phi ne}}
	\end{pmatrix}_{(\Lambda)}.
	\label{eq:adm_c5}
	\eeq
	
	To study the running numerically, we set $\{prst\} = \{1111\}$ for illustration. We list the $16\times 16$ ADM in the basis 
	\beqa
	\vec{\mathbf{\C}} &=&\{\C_{nd},\,\C_{nu},\,\C_{ne},\,\C_{qn},\,\C_{\ell n},\,\C_{\phi n},\,\C_{n\phi},\,\C_{nW},\,\C_{nB},\,C^{(1)}_{\ell nqd},\,C^{(3)}_{\ell nqd},\,\C_{nedu},\,\C_{\ell n \ell e},\,\C_{\ell nuq},\,\C_{\phi ne},\,\C_{nn}\}\,.\notag\\
	\eeqa
	The gauge couplings at 1~TeV are set to $g_1 = 0.36,\, g_2 = 0.64,\, g_3 = 1.1$~\cite{Zyla:2020zbs}.
	The 16 WCs at $M_Z$ and at $\Lambda =1$~TeV are related by
	\\
	
	\beq
	\frac{\delta \mathbf{\C}(M_Z)}{10^{-3}} =
	\begin{psmallmatrix}
		-0.87 & 1.7 & -0.87 & 0.87 & -0.87 & 0.43  & & & & & & & & & & \\
		1.7 & -3.5 & 1.7 & -1.7 & 1.7 & -0.87 &  & & & & & & & & & \\ 
		-2.6 & 5.2 & -2.6 & 2.6 & - 2.6 & 1.3 & & & & & & & & & & \\ 
		0.44 & -0.87 & 0.44 & -0.44 & 0.44 & -0.22 & & & & & & & &  & & \\ 
		-1.3 & 2.6 & -1.3 & 1.3 & -1.3 & 0.65 & & & & & & & & & & \\ 
		1.3 & -2.6 & 3.9 & -1.3 & 1.3 & -0.65 & & & & & & & & & & \\ 
		& & & & &  & 46 & 39 & -8.8 & & & & & & & & \\ 
		& & & & &  & 0 & 7.1 & 5.2 & & & & & &  &\\ 
		& & & & &  & 0 & 16 & -0.99 & & & & & &  &\\ 
		& & & & &  &  & &  & 150 & -110 & & & &  &\\ 
		& & & & &  &  & & & -2.2 & -29 & & & &  &\\
		& & & & &  &  & & &  & & 7.9 & & &  &\\
		& & & & &  &  & & &  & & & -0.43 & &  &\\
		& & & & &  &  & & &  & & &  & 150 &  &\\
		& & & & &  &  & & &  & & &  & & 5.9 &\\
		& & & & &  &  & & &  & & &  & & & 0\\
	\end{psmallmatrix} \mathbf{\C}(\Lambda)\,.
	\eeq
	\\
	
	The running effects in the $6\times 6$ and $3\times 3$ blocks are small because only electroweak gauge couplings contribute. The mixing in the $2\times 2$ block is large as it is governed by QCD.
	
	\section{Phenomenology}
	\label{sec:pheno}
	We briefly comment on some phenomenological consequences of our results; for earlier work see Refs.~\cite{Alcaide:2019pnf, Butterworth:2019iff, Chala:2020vqp, Li:2020lba, Biekotter:2020tbd, Li:2020wxi}. Semileptonic decays of the $b$ quark are topical
	given that both charged current and neutral current decay measurements are hinting at new physics. SMNEFT operators lead to the
	charged current decay $ b \to c \ell \bar{n}$, which contributes  at the hadronic level to $B \to  D^{(*)} \tau \bar{\nu}_\tau$.  They also generate the neutral current decay $ b \to s \bar{n} n$ which contributes  at the hadronic level to $ B \to K^{(*)} + \text{invisible}$ decays, which is interpreted as $ B \to K^{(*)} \bar{\nu} \nu$ in the SM.  
	In the lepton sector, of interest are the FCNC decays $ \tau \to \mu + \text{invisible}$ and $ \mu \to e + \text{invisible}$. To make contact with  low-energy phenomenology, we first run the RG equations down to the weak scale and then match to the low-energy effective field theory extended with right-handed neutrinos $n$ (LNEFT). Depending on the process, further RG running must be performed from the electroweak scale to the appropriate low energy scale such as the
	$m_b$ scale for $B$ meson decay and the $m_\tau$ scale for $\tau$ decay.
	Note that the sterile neutrino can mix with the active neutrinos, which in itself produces interesting  phenomenology, but to keep our discussion simple we neglect this mixing. We select the following four types of process and list the SMNEFT operators relevant to them:
	\begin{itemize}
		\item{ \text{$B \to  D^{(*)} \tau \bar{\nu}_\tau$: $\Q_{nedu}$, $\Q_{\ell nuq}$, $\Q^{(1)}_{\ell nqd}$,  and  $\Q^{(3)}_{\ell nqd}$}}
		\item {\text{$B\to K^{(*)} \nu \bar \nu$ {\rm \&} $K\to \pi \nu \bar \nu$: $\Q_{nd}$, $\Q_{qn}$, $\Q^{(1)}_{\ell nqd}$, and  $\Q^{(3)}_{\ell nqd}$}}
		\item {\text{$t\to c \nu \bar \nu$ \& $c\to u \nu \bar \nu$: $\Q_{nu}$, $\Q_{qn}$, and $\Q_{\ell nuq}$}}
		\item { {\text{$\tau \to \mu \nu \bar \nu$ \& $\mu \to e \nu \bar \nu$: $\Q_{ne}$, $\Q_{\ell n}$, and $\Q_{\ell n\ell e}$}}}
	\end{itemize}
	The FCNC operators, $\Q_{nd}$, $\Q_{nu}$, $\Q_{ne}$, $\Q_{qn}$ and $\Q_{\ell n}$ do not run when only gauge interactions are considered. So we do not study these operators and focus on the five operators, $\Q_{nedu}$, $\Q_{\ell nuq}$, $\Q^{(1)}_{\ell nqd}$, $\Q^{(3)}_{\ell nqd}$ and $\Q_{\ell n\ell e}$. Interestingly, $\Q_{\ell nuq}$, $\Q^{(1)}_{\ell nqd}$ and  $\Q^{(3)}_{\ell nqd}$ can contribute to both the charged current  and neutral current decays, and to coherent elastic neutrino-nucleus scattering~\cite{Han:2020pff}.
	For certain flavor combinations, $\Q_{\ell n\ell e}$ can produce both $\tau \to \mu$ and $\mu \to e$ decays.

	Before studying the low-energy phenomenology, we first run the operators down from  the new physics scale $\Lambda$ to the weak scale $\mu_{EW}$. By using the leading-log approximation in Eq.~(\ref{eq:ll}), we relate the values of the WCs at $M_Z$  to  their values at  1~TeV:
	\beqa
	\begin{pmatrix}
		\C_{\us{prst}{nedu}}\\
		\C_{\us{prst}{\ell nuq}}\\
		C^{(1)}_{\us{prst}{\ell nqd}}\\
		C^{(3)}_{\us{prst}{\ell nqd}}
	\end{pmatrix}_{(M_Z)} &=& \begin{pmatrix}
		1.0 & 0 & 0 & 0\\
		0 & 1.1 &  0 & 0\\
		0 &  0 & 1.1 & -0.11\\
		0 & 0 & -2.2\times 10^{-3} & 0.97 \\
	\end{pmatrix} \begin{pmatrix}
		\C_{\us{prst}{nedu}}\\
		\C_{\us{prst}{\ell nuq}}\\
		C^{(1)}_{\us{prst}{\ell nqd}}\\
		C^{(3)}_{\us{prst}{\ell nqd}}
	\end{pmatrix}_{(\text{1 TeV})},\\
	\begin{pmatrix}
		\C_{\underset{prst}{\ell n\ell e}}\\
		\C_{\underset{srpt}{\ell n\ell e}}\\
	\end{pmatrix}_{(M_Z)} &=& \begin{pmatrix}
		1.0 & -1.3\times 10^{-2} \\
		-1.3\times 10^{-2} & 1.0\\
	\end{pmatrix} \begin{pmatrix}
		\C_{\underset{prst}{\ell n\ell e}}\\
		\C_{\underset{srpt}{\ell n\ell e}}\\
	\end{pmatrix}_{(\text{1 TeV})}.
	\eeqa
	
	To study the phenomenology at energies below the electroweak scale one can no longer use SMNEFT  because of  electroweak symmetry breaking.  Instead,  LNEFT,  which respects the ${SU(3)}_C \times U(1)_Q$ symmetry must be employed to study the processes listed above. We introduce the relevant  LNEFT operators and match them with the SMNEFT operators at the weak scale. The SMNEFT operators can generate both neutral and charged current processes after electroweak symmetry breaking. The induced LNEFT operators in the convention of Ref.~\cite{Bischer:2019ttk} are displayed in Table~\ref{table:match} and their matching relations at tree level are
	\beq
	\begin{split}
		C^{V,RR}_{\underset{prst}{nedu}}&=\C_{\underset{prst}{nedu}}\,,\quad C^{S,RL}_{\underset{prst}{en ud}}=\C_{\underset{prst}{\ell nuq}}\,,\quad C^{S,RR}_{\underset{prst}{enud}}=-C^{(1)}_{\underset{pr\delta t}{\ell nqd}} \frac{V_{s\delta}}{V_{st}}\,,\\
		\quad C^{T,RR}_{\underset{prst}{enud}}&=-C^{(3)}_{\underset{pr\delta t}{\ell nqd}}\frac{V_{s\delta}}{V_{st}}\,,\quad C^{S,RR}_{\underset{prst}{en\nu e}} =- \C_{\underset{prst}{\ell n\ell e}}\,,
	\end{split}
	\eeq
	
	\beq
	C^{S,RL}_{\underset{prst}{\nu nuu}}=\C_{\underset{prs\delta}{\ell nuq}} V^*_{t\delta}\,,\quad C^{S,RR}_{\underset{prst}{\nu ndd}}=C^{(1)}_{\underset{prst}{\ell nqd}}\,,\quad C^{T,RR}_{\underset{prst}{\nu ndd}}=C^{(3)}_{\underset{prst}{\ell nqd}}\,,\quad C^{S,RR}_{\underset{prst}{\nu nee}} = \C_{\underset{prst}{\ell n\ell e}}\,.
	\eeq
	where we chose a flavor basis in which the left-handed down-type quarks and charged leptons are aligned. The flavor basis for up-type quarks in terms of the mass basis is given by $V^\dagger u_L$,
	where $V$ is the SM CKM matrix. The neutrino fields are in the flavor basis for convenience.
	\begin{table}
		\begin{center}
			\renewcommand{\arraystretch}{2.}
			\begin{tabular}{ |c|c|c| } 
				\hline
				SMNEFT & NC LNEFT & CC LNEFT \\ 
				\hline
				$\Q_{\underset{prst}{nedu}}$ & - & $O^{V,RR}_{\underset{prst}{nedu}} = (\bar{n}_{Rp}\gamma^\mu e_{Rr}) (\bar{d}_{Rs}\gamma^\mu u_{Rt} )$  \\
				\hline 
				$\Q_{\underset{prst}{\ell nuq}}$ & $O^{S,RL}_{\underset{prst}{\nu nuu}} = (\bar{\nu}_{Lp} n_{Rr}) (\bar{u}_{Rs} u_{Lt} )$  & $O^{S,RL}_{\underset{prst}{enud}} = (\bar{e}_{Lp} n_{Rr}) (\bar{u}_{Rs} d_{Lt} )$ \\ 
				\hline
				$\Q^{(1)}_{\underset{prst}{\ell nqd}}$ &  $O^{S,RR}_{\underset{prst}{\nu ndd}} = (\bar{\nu}_{Lp} n_{Rr}) (\bar{d}_{Ls} d_{Rt} )$   & $O^{S,RR}_{\underset{prst}{en ud}} = (\bar{e}_{Lp} n_{Rr}) (\bar{u}_{Ls} d_{Rt} )$ \\ 
				\hline
				$\Q^{(3)}_{\underset{prst}{\ell nqd}}$ & $O^{T,RR}_{\underset{prst}{\nu ndd}} = (\bar{\nu}_{Lp}\sigma^{\mu\nu} n_{Rr}) (\bar{d}_{Ls}\sigma_{\mu\nu} d_{Rt} )$ & $O^{T,RR}_{\underset{prst}{en ud}} = (\bar{e}_{Lp} \sigma^{\mu\nu}n_{Rr}) (\bar{u}_{Ls}\sigma_{\mu\nu} d_{Rt} )$ \\ 
				\hline
				$\Q_{\underset{prst}{\ell n\ell e}}$ & $O^{S,RR}_{\underset{prst}{\nu nee}} = (\bar{\nu}_{Lp} n_{Rr}) (\bar{e}_{Ls} e_{Rt} )$ & $O^{S,RR}_{\underset{prst}{en\nu e}} = (\bar{e}_{Lp} n_{Rr}) (\bar{\nu}_{Ls} e_{Rt} )$ \\ 
				\hline
			\end{tabular}
		\end{center}
		\caption{Operator structure matching between SMNEFT and LNEFT.}
		\label{table:match}
	\end{table}
	%With the running and matching effects above the weak scale,
	In the next subsections, we study the low-energy phenomenology of the listed processes.

	\subsection{\boldmath{$B \to  D^{(*)} \tau \bar{\nu}$}}
	The CC LNEFT operators induced by the SMNEFT operators $\Q_{\underset{\alpha332}{nedu}},  \Q_{\underset{3\alpha23}{\ell nuq}}, \Q_{\underset{3\alpha23}{\ell nqd}}^{(1)}\,\text{and}~\Q_{\underset{3\alpha23}{\ell nqd}}^{(3)}$ can affect this process; see Table~\ref{table:match}. Here, $\alpha$ is the flavor index of the right-handed neutrino $n$.
	Accounting for QED and QCD running below the weak scale, the one-loop RGE for the four LNEFT operators is given by
	%\beqa
	%\mu\frac{d}{d\mu} C^{V,RR}_{\underset{prst}{\nu edu}} &=& (Q_d^2+(Q_e+Q_u)^2-2Q_d(Q_e+Q_u)+6Q_eQ_u) \frac{\alpha_e}{4\pi} C^{V,RR}_{\underset{prst}{\nu edu}},\\
	%\mu\frac{d}{d\mu} C^{S,RL}_{\underset{prst}{e\nu ud}} &=& [(Q_d^2+(Q_e+Q_u)^2-2Q_d(Q_e+4Q_u)) \frac{\alpha_e}{4\pi} - 8 \frac{\alpha_3}{4 \pi}] C^{V,RR}_{\underset{prst}{e\nu ud}},\\
	%\mu\frac{d}{d\mu} C^{S,RR}_{\underset{prst}{e\nu ud}} &=& [(Q_d^2+(Q_e+Q_u)^2-2Q_d(Q_e+4Q_u)) \frac{\alpha_e}{4\pi} - 8 \frac{\alpha_3}{4 \pi}] C^{S,RR}_{\underset{prst}{e\nu ud}},\\
	%&-& 24Q_e(Q_d+Q_u) \frac{\alpha_e}{4 \pi}C^{T,RR}_{\underset{prst}{e\nu ud}},\\
	%\mu\frac{d}{d\mu} C^{T,RR}_{\underset{prst}{e\nu ud}} &=& -\frac{1}{2}Q_e(Q_e+2Q_u)\frac{\alpha_e}{4\pi} %&+& [2(Q_u^2+Q_eQ_u-2Q_e^2) \frac{\alpha_e}{4 \pi} + \frac{8}{3}\frac{\alpha_3}{4\pi}]C^{T,RR}_{\underset{prst}{e\nu ud}}.
	%\eeqa
	\beq
	\begin{pmatrix}
		\dot C^{V,RR}_{\us{\alpha332}{nedu}}\\
		\dot C^{S,RL}_{\us{3\alpha23}{enud}}\\
		\dot C^{S,RR}_{\us{3\alpha23}{enud}}\\
		\dot C^{T,RR}_{\us{3\alpha23}{enud}}\\
	\end{pmatrix}_{(\mu)} = [e^2\begin{pmatrix}
		-4 & 0 & 0 & 0 \\
		0 & \frac{4}{3} & 0 & 0\\
		0 &  0 & \frac{4}{3} & 8\\
		0 & 0 & \frac{1}{6} & -\frac{40}{9} \\
	\end{pmatrix}
	+g_3^2\begin{pmatrix}
		0 & 0 & 0 & 0\\
		0 & -8 &  0 & 0\\
		0 &  0 & -8 & 0\\
		0 & 0 & 0 & \frac{8}{3} \\
	\end{pmatrix}] \begin{pmatrix}
		C^{V,RR}_{\us{\alpha332}{nedu}}\\
		C^{S,RL}_{\us{3\alpha23}{enud}}\\
		C^{S,RR}_{\us{3\alpha23}{enud}}\\
		C^{T,RR}_{\us{3\alpha23}{enud}}\\
	\end{pmatrix}_{(\mu)},
	\label{eq:adm_eud}
	\eeq
	where $e$ is the QED coupling. Using Eq.~(\ref{eq:ll}), we relate the four LNEFT operators at the $m_b$  and $M_Z$ scales:
	\beq
	\begin{pmatrix}
		C^{V,RR}_{\us{\alpha332}{nedu}}\\
		C^{S,RL}_{\us{3\alpha23}{enud}}\\
		C^{S,RR}_{\us{3\alpha23}{enud}}\\
		C^{T,RR}_{\us{3\alpha23}{enud}}\\
	\end{pmatrix}_{(m_b)} = \begin{pmatrix}
		1.0 & 0 & 0 & 0\\
		0 & 1.2 &  0 & 0\\
		0 &  0 & 1.2 & -1.5\times10^{-2}\\
		0 & 0 & -3.2\times 10^{-4} & 0.93 \\
	\end{pmatrix} \begin{pmatrix}
		C^{V,RR}_{\us{\alpha332}{nedu}}\\
		C^{S,RL}_{\us{3\alpha23}{enud}}\\
		C^{S,RR}_{\us{3\alpha23}{enud}}\\
		C^{T,RR}_{\us{3\alpha23}{enud}}\\
	\end{pmatrix}_{(M_Z)}.
	\eeq
	The mixing between $O^{S,RR}_{enud}$ and $O^{T,RR}_{enud}$ is small as it is induced by QED. However, the corresponding mixing of the SMNEFT operators is relatively strong as it comes from electroweak effects. 
	
	\subsection{\boldmath{$B\to K^{(*)} \nu \bar \nu$  \& $ K \to \pi \nu \bar \nu $}}
	$B\to K^{(*)} + \text{invisible}$ decay, which would be interpreted as $B\to K^{(*)} \nu \bar \nu$ in the SM, is produced by $O^{S,RR}_{\nu ndd}$ and $O^{T,RR}_{\nu ndd}$. The flavor structures are $\{prst\} = \{\alpha\beta23\}$. 
	The process $ K \to \pi  \nu \bar \nu$ can also be generated with the flavor structures, $\{prst\} = \{\alpha\beta12\}$. 
	%In many scenarios,  new physics directly affects only the third generation in the flavor basis and other processes are generated via quark mixing~\cite{Bhattacharya:2014wla,Bhattacharya:2016mcc}. In  this scenario one would start with the operator $ b \bar{b} n \bar{n} $ and assuming the down quark mixing to be hierarchical, at leading order, the dominant  FCNC decay is $ b \to s n \bar{n}$.
	The ADM for $O^{S,RR}_{\nu ndd}$ and $O^{T,RR}_{\nu ndd}$ is
	\beq
	\begin{pmatrix}
		\dot C^{S,RR}_{\us{\alpha\beta23}{\nu ndd}}\\
		\dot C^{T,RR}_{\us{\alpha\beta23}{\nu ndd}}\\
	\end{pmatrix}_{(\mu)} = [e^2\begin{pmatrix}
		-\frac{2}{3} & 0  \\
		0 & \frac{2}{9} \\
	\end{pmatrix}
	+g_3^2\begin{pmatrix}
		-8 & 0\\
		0 & \frac{8}{3}  \\
	\end{pmatrix}] \begin{pmatrix}
		C^{S,RR}_{\us{\alpha\beta23}{\nu ndd}}\\
		C^{T,RR}_{\us{\alpha\beta23}{\nu ndd}}\\
	\end{pmatrix}_{(\mu)}.
	\label{eq:adm_nd}
	\eeq
	The WCs at $m_b$ and $M_Z$ are related by
	\beq
	\begin{pmatrix}
		C^{S,RR}_{\us{\alpha\beta23}{\nu ndd}}\\
		C^{T,RR}_{\us{\alpha\beta23}{\nu ndd}}\\
	\end{pmatrix}_{(m_b)} =\begin{pmatrix}
		1.2 & 0  \\
		0 & 0.92\\
	\end{pmatrix}
	\begin{pmatrix}
		C^{S,RR}_{\us{\alpha\beta23}{\nu ndd}}\\
		C^{T,RR}_{\us{\alpha\beta23}{\nu ndd}}\\
	\end{pmatrix}_{(M_Z)}.
	\eeq
	While there is no mixing between the NC LNEFT operators, their corresponding SMNEFT operators can mix above the weak scale. For $ K \to \pi  \nu \bar \nu$  one has to run down to a scale appropriate for kaon decays.
	
	\subsection{\boldmath{$ t \to c \nu \bar{\nu}$  \&  $c \to u \nu \bar{\nu} $}}
	The NC LNEFT operator $O^{S,RL}_{\nu nuu}$ induced by $\Q_{\ell nuq}$ can generate the rare decay $ t \to c \nu \bar{\nu}$ with $\{prst\} = \{\alpha\beta23\}$. The RG equation for $O^{S,RL}_{\nu nuu}$ below the weak scale is
	\beq
	\dot C^{S,RL}_{\nu nuu}(\mu) = [e^2(-\frac{8}{3}) +  g_3^2(-8)]C^{S,RL}_{\nu nuu}(\mu)\,,
	\eeq
	and
	\beq
	C^{S,RL}_{\nu nuu}(\mu=m_b) = 1.2~C^{S,RL}_{\nu nuu}(\mu = M_Z)\,.
	\eeq
	
	\subsection{ \boldmath{$\tau \to \mu \nu \bar \nu$ \rm \& $\mu \to e \nu \bar \nu$}}
	The decays
	$\tau \to \mu + \text{invisible}$ and $\mu \to e + \text{invisible}$ are generated by $O^{S,RR}_{\nu nee}$ and $O^{S,RR}_{en\nu e}$.  Note that the flavor is mixed for $\Q_{\ell n\ell e}$.  The flavor combination $\{prst\} = \{1132\}$ can generate both  $\tau \to \mu$ and $\mu \to e$ decays. The relevant operators are 
	$O^{S,RR}_{\underset{1132}{\nu nee}}$, $O^{S,RR}_{\underset{1132}{en\nu e}}$, $O^{S,RR}_{\underset{3112}{\nu nee}}$ and $O^{S,RR}_{\underset{3112}{en\nu e}}$. The running at one-loop order is given by
	\beq
	\begin{pmatrix}
		\dot C^{S,RR}_{\underset{1132}{\nu nee}}\\
		\dot C^{S,RR}_{\underset{3112}{en\nu e}}\\
		\dot C^{S,RR}_{\underset{1132}{en\nu e}}\\
		\dot C^{S,RR}_{\underset{3112}{\nu nee}}\\
	\end{pmatrix}_{(\mu)} = e^2\begin{pmatrix}
		-6 & 4 &  &  \\
		0 & 2 &  & \\
		&  & -6 & 4\\
		&  & 0 & 2\\
	\end{pmatrix} \begin{pmatrix}
		C^{S,RR}_{\underset{1132}{\nu e}}\\
		C^{S,RR}_{\underset{3112}{e\nu\nu e}}\\
		C^{S,RR}_{\underset{1132}{e\nu\nu e}}\\
		C^{S,RR}_{\underset{3112}{\nu e}}\\
	\end{pmatrix}_{(\mu)}.
	\label{eq:adm_ne}
	\eeq
	The WCs  at $m_\tau$ and $M_Z$ are related by
	\beq
	\begin{pmatrix}
		C^{S,RR}_{\underset{1132}{\nu nee}}\\
		C^{S,RR}_{\underset{3112}{en\nu e}}\\
		C^{S,RR}_{\underset{1132}{en\nu e}}\\
		C^{S,RR}_{\underset{3112}{\nu nee}}
	\end{pmatrix}_{(m_\tau)} = \begin{pmatrix}
		1.0 & -9.9\times 10^{-3} &  &  \\
		0 & 1.0 &  & \\
		& &1.0 & -9.9\times 10^{-3}   \\
		& & 0 & 1.0 \\
	\end{pmatrix} \begin{pmatrix}
		C^{S,RR}_{\underset{1132}{\nu nee}}\\
		C^{S,RR}_{\underset{3112}{en\nu e}}\\
		C^{S,RR}_{\underset{1132}{en\nu e}}\\
		C^{S,RR}_{\underset{3112}{\nu nee}}
	\end{pmatrix}_{(M_Z)}.
	\eeq
	The small mixing between these operators is a consequence of QED. For muon decay, one needs to run down to the muon mass.

	\subsection{\bf Electroweak precision observables}
	The operators $\mathcal{O}_{\phi n}$ and $\mathcal{O}_{\phi n e}$ give rise to RH $Z$-couplings to $n$ and RH $W$ couplings to $n$ and leptons.
	The RH $Z$ couplings to $n$ can be parameterized in terms of the Wilson coefficient
	$\mathcal{C}_{\phi n}$ as
	\begin{equation}
	\delta \mathcal{L}_Z = -\frac{g_Z }{2} v^2 [\mathcal{C}_{\phi n}]_{pr}  ~(\bar n_p \gamma_\mu n_r)~ Z_\mu\,,
	\end{equation}
	where $g_Z^2= g_1^2+g_2^2$.
	Therefore, $\mathcal{C}_{\phi n}$ contributes to the $Z$-width via
	$\Gamma(Z \to n \bar n)$.
	Similary, the RH $W$ couplings can be parameterized in terms of $\mathcal{C}_{\phi ne}$ as
	\begin{equation}
	\delta \mathcal{L}_W = -\frac{g_2}{2 \sqrt 2} v^2 [\mathcal{C}_{\phi ne}]_{pr}
	~(\bar n_p \gamma^\mu e_r)  ~W_\mu^+ + h.c.\,.
	\end{equation}
	Note that such leptonic RH $W$ couplings are absent in SMEFT because the
	RH neutrino field is absent.
	The modified $Z$ and $W$ couplings affect electroweak precision observables.
	Interestingly, while $\mathcal{O}_{\phi ne}$ does not mix with the other operators as can be seen from Eq.~\eqref{eq:adm_phine},
	$\mathcal{O}_{\phi  n}$  has mixing with other operators; see Eq.~\eqref{eq:adm_phin}. Hence,
	electroweak precision observables can place indirect
	constraints on the $\mathcal{O}_{nd}$, $\mathcal{O}_{nu}$, $\mathcal{O}_{ne}$,
	$\mathcal{O}_{qn}$ and $\mathcal{O}_{\ell n}$ operators that mix with $\mathcal{O}_{\phi  n}$, by a global fit.

	\section{Summary}
	\label{sec:sum}
	We presented the gauge terms  of the one-loop anomalous dimension matrix for the dimension-six operators of SMNEFT; see Eqs.~(\ref{eq:adm_c1}) to (\ref{eq:adm_c5}).  
	We found that renormalization group evolution introduces interesting correlations among observables in different sectors. 
	We discussed a few phenomenological implications of our results. To make contact with low energy observables we  also included the matching of SMNEFT to LNEFT at the weak scale and RGE below the weak scale.
	However, to be confident that cancellations of terms between
	independent operators are absent, the full one-loop RGE must be calculated. 
	%. The mixing the matching relations between LNEFT operatos can  introduce nontrivially correlations. To set the bounds on SMNEFT operators, we have to consistently running down %to the corresponding physical scale.

	\acknowledgments
	This work was supported by NSF Grant No. PHY1915142 (A.D.), NSERC of Canada (J.K.), DOE Grant No. DE-FG02-95ER40896 and PITT PACC (H.L.), and DOE Grant No. de-sc0010504 (D.M.).

	\begin{appendix}
		\section{Derivation of anomalous dimensions}
		\label{app:derivations}
		%%%%%%%%%%%%%%%%%%%%%%%%%%%%%%%%%%%%%%%%%%%%%%%%
		
		%ijkl
		\subsection{Master formulae for the ADM}
		\label{app:master}	
		\noindent	
		In this appendix, we collect the master formulae \cite{Buras:1998raa} 
		for computing the ADM in the context of low energy effective field theory 
		due to one-loop QCD corrections and generalize them to the electroweak interactions 
		and use them for deriving ADMs in SMNEFT. 
		Consider an insertion of a four-fermion $\psi^4$ operator 
		$(\bar \psi_1^{\alpha,i} (\hat V_{1})_{\alpha\beta, ij} \Gamma_{1} \psi_2^{\beta,j}) \otimes
		(\bar \psi_3^{\gamma,k} (\hat V_{2})_{\gamma \delta, kl} \Gamma_2 \psi_4^{\delta,l})$ 
		into the diagrams in Fig.~\ref{fig:current-current} (current-current topology) and 
		\ref{fig:penguin}(d) (penguin topology). Here, $\hat V_1 \otimes \hat V_2$ represent the
		$SU(3)_C$ and $SU(2)_L$ structure of the operator, and the color and isospin 
		indices are denoted by Greek and Latin letters.	
		For the four-fermion operators in Table~\ref{tab:SMNEFTops}, $\hat V_1 \otimes \hat V_2 $ are 	
		$\delta_{\alpha \beta}$, $\delta_{ij}$, 
		$\delta_{\alpha \beta} \delta_{ij}$,
		$\varepsilon_{jk}$, 
		$\delta_{\alpha \beta} \varepsilon_{jk}$, etc., and $\Gamma_{1,2}$ are Dirac matrices.
		
		The UV divergent parts of the current-current diagrams in Fig.~\ref{fig:current-current} 
		mediated by the gauge boson $X_\mu= (B_\mu, W_\mu, G_\mu)$ depend on $\alpha_m$ and are given by
		\\
		\begin{eqnarray}
		\label{eq:div:cc}
		\mathcal{D}_a  &=& \mathcal{D}_a^{(1)}+ \mathcal{D}_a^{(2)} =
		\frac{\alpha_m}{4\pi} \frac{1}{4\varepsilon}
		\left (\mathcal{C}_a^{(1)} \gamma_\mu \gamma_\rho \Gamma_1 \gamma^\rho \gamma^\mu
		\otimes \Gamma_2  +  
		\mathcal{C}_a^{(2)} \Gamma_1 \otimes \gamma_\mu \gamma_\rho \Gamma_2 \gamma^\rho \gamma^\mu
		\right )\,,\\
		\label{eq:div:cc2}
		\mathcal{D}_b  &=&  \mathcal{D}_b^{(1)}+ \mathcal{D}_b^{(2)} =
		-\frac{\alpha_m}{4\pi} \frac{1}{4\varepsilon}
		\left (\mathcal{C}_b^{(1)} \Gamma_1 \gamma_\rho \gamma_\mu \otimes 
		\Gamma_2 \gamma^\rho \gamma^\mu +
		\mathcal{C}_b^{(2)} \gamma_\mu \gamma_\rho \Gamma_1 \otimes \gamma^\mu \gamma^\rho \Gamma_2 \right )\,,\\
		\label{eq:div:cc3}
		\mathcal{D}_c  &=&  \mathcal{D}_c^{(1)}+ \mathcal{D}_c^{(2)} =
		\frac{\alpha_m}{4\pi} \frac{1}{4\varepsilon}
		\left (\mathcal{C}_c^{(1)} \Gamma_1 \gamma_\rho \gamma_\mu \otimes
		\gamma^\mu \gamma^\rho \Gamma_2  +
		\mathcal{C}_c^{(2)} \gamma_\mu \gamma_\rho \Gamma_1 \otimes \Gamma_2 \gamma^\rho \gamma^\mu \right )\,.
		\end{eqnarray}
		In dimensional regularization, we use the convention $d = 4 - 2\varepsilon$. 
		Here, $\mathcal{D}_{a,b,c}^{(1)}$ represent the symmetric counterparts of the diagrams 
		$\mathcal{D}_{a,b,c}^{(2)}$ shown in Fig.~\ref{fig:current-current}. The two terms in 
		Eqs.~\eqref{eq:div:cc}-\eqref{eq:div:cc3} correspond to these two kinds of diagrams.
		The coefficients are given by
		\\
		\begin{eqnarray}
		\mathcal{C}_a^{(1)} &=& J_m^x \hat V_1 J_m^x \otimes \hat V_2\,, \quad
		\mathcal{C}_a^{(2)} = \hat V_1 \otimes J_m^x \hat V_2 J_m^x\,,   \\
		\mathcal{C}_b^{(1)} &=& \hat V_1 J_m^x  \otimes \hat V_2 J_m^x\,, \quad
		\mathcal{C}_b^{(2)} = J_m^x \hat V_1 \otimes J_m^x \hat V_2 \,,   \\
		\mathcal{C}_c^{(1)} &=&  \hat V_1 J_m^x \otimes J_m^x \hat V_2  \,,   \quad
		\mathcal{C}_c^{(2)} = J_m^x \hat V_1 \otimes \hat V_2 J_m^x\,, 
		\end{eqnarray}
		where $J_m^x$ $(m=1,2,3)$ are the $SU(3)_C$, $SU(2)_L$ and $U(1)_Y$  
		generators.
		The sum over the index $x$ is implied.
		
		For the penguin insertion in Fig.~\ref{fig:penguin}(d), the UV divergent 
		part, if we close $\hat V_1 \Gamma_1$ part of the inserted operator, is given by 
		\\
		\begin{equation}
		\label{eq:div:penguin}
		\mathcal{D}_d  =
		\mathcal{C}_d \frac{\alpha_m}{4\pi} \frac{1}{4\varepsilon}
		\left [\frac{1}{6} \frac{q^\mu q^\nu}{q^2} + \frac{g^{\mu \nu}}{12}  \right ]
		{\rm Tr} (\Gamma_1 \gamma_\mu \gamma_\lambda \gamma_\nu) \Gamma_2 
		\otimes \gamma^\lambda\,,
		\end{equation}
		with the coefficient $\mathcal{C}_d = {\rm Tr}(\hat V_1 J_m^x) \hat V_2 \otimes J_m^x $. 
		Note that, depending upon the structure of the operator given by the matrix ($\hat V_{1}$) and 
		the type of gauge boson mediated in the penguin diagram Fig.~\ref{fig:penguin}(d), 
		the trace can be over the $SU(3)_C$ or $SU(2)_L$ indices.

		\subsection{Field strength renormalization}
		\noindent	
		The field strength renormalization constants are defined in 
		Eq.~\eqref{eq:constants:fields}. At one-loop, these are given by the
		coefficients
		\\
		\begin{eqnarray}
		\label{eq:field-constants}
		a_{q,u,d}^3 &=& -\frac{4}{3} \,, \quad a^3_{\ell, e} = 0\,,   \\
		a^2_{q,\ell} &=& -\frac{3}{4} \,, \quad a^2_{u,d,e}=0\,, \\
		a^1_\psi &=& -y_{\psi}^2\,,  \quad a_n^{1,2,3}=0\,,
		\end{eqnarray}
		where $y_\psi$ is the hypercharge of the 
		fields $\psi= \{q, u, d, \ell, e\}$. \\
		
		\subsection{Operator renormalization}
		\noindent	
		
		For illustration, we present an explicit computation of the 
		renormalization constants for $\mathcal{O}_{nd}$, 
		$\mathcal{O}_{nedu}$, $\mathcal{O}_{\ell n \ell e}$ and $\mathcal{O}_{\phi n}$.
		For the other operators, a similar procedure can be followed. Here we present $\gamma$, and in section~\ref{sec:results} we present $\gamma_\C = \gamma^T$.

		\begin{boldmath}
			\subsubsection{$\mathcal{O}_{nd}$-$\mathcal{O}_{nd}$ mixing}
		\end{boldmath}
		\noindent	
		To extract the divergent pieces of the diagrams we use the master formulae 
		of Appendix~\ref{app:master}. For the insertion of
		$\mathcal{O}_{nd} =(\bar n_p \gamma_\mu n_r)(\bar d_s \gamma^\mu d_t)$ to 
		generate the same, we have 
		\\
		\begin{eqnarray}
		\psi_1 &=& n_p\,, \quad \psi_2= n_r\,,  \quad \psi_3 = d_s\,,  \quad \psi_4= d_t\,, \\
		\hat V_1 &\otimes& \hat{V_2} = \delta_{\alpha \beta} \,, \quad 
		\Gamma_1 \otimes \Gamma_2 = \gamma_\mu P_R \otimes \gamma^\mu P_R\,.
		\end{eqnarray}
		In this case, $\mathcal{D}_a^{(2)}$, i.e. the
		first topology in Fig.~\ref{fig:current-current} with 
		$X_\mu= G_\mu$ or $B_\mu$ connected between two $d$-quark legs is generated. 
		Using Eq.~(\ref{eq:div:cc}), the divergent parts of these two contributions are given by
		\\
		\begin{eqnarray}
		\mathcal{D}_a^{(2)}[B] &=&  \frac{\alpha_1}{4\pi} \frac{1}{4} \frac{1}{\varepsilon} (4y_d^2) (\gamma_\mu P_R \otimes \gamma^\mu P_R) \,,\\
		\mathcal{D}_a^{(2)}[G] &=& \frac{\alpha_3}{4 \pi} \frac{1}{4}\frac{1}{\varepsilon} {16 \over 3} (\gamma_\mu P_R \otimes \gamma^\mu P_R)\,.
		\end{eqnarray}\\
		
		\begin{boldmath}
			\subsubsection{$\mathcal{O}_{nd}$-$\mathcal{O}_{nu}$ mixing}
		\end{boldmath}
		\noindent	
		Next, we consider the penguin insertion which leads to $\mathcal{O}_{nd}$-$\mathcal{O}_{nu}$
		operator mixing through $\mathcal{D}_d[B]$. The penguin insertion 
		of $\mathcal{O}_{nu}$ leads to operator $\mathcal{O}_{nd}$ for
		\begin{eqnarray}
		\psi_1 &=& n_p\,, \quad \psi_2= n_r\,,  \quad \psi_3 = d_s\,,  \quad \psi_4= d_t\,, \\
		\hat V_1 &\otimes& \hat{V_2} = \delta_{\alpha \beta}\,, \quad 
		\Gamma_1 \otimes \Gamma_2 = \gamma_\mu P_R \otimes \gamma^\mu P_R\,.
		\end{eqnarray}
		Now, using the Eq.~(\ref{eq:div:penguin}), we obtain
		\begin{equation}
		\mathcal {D}_d [B] = -\frac{\alpha_1}{4\pi}\frac{1}{\varepsilon} 
		\frac{2}{3} y_d y_u N_c(\gamma_\mu P_R \otimes \gamma^\mu P_R)\,.
		\end{equation}\\
		
		\begin{boldmath}
			\subsubsection{$\mathcal{O}_{nd}$-$\mathcal{O}_{\phi n}$ mixing}
		\end{boldmath}
		\noindent
		As an example of fermionic and bosonic oprator mixing, we present the mixing between $\mathcal{O}_{nd}$ and $\mathcal{O}_{\phi n}$, which is given by 
		Fig.~\ref{fig:penguin}(e). Its divergent part reads
		\begin{equation}
		\mathcal{D}_e[B] = -\frac{\alpha_1}{4\pi}\frac{1}{\varepsilon} 
		\frac{2}{3} y_d y_h (\gamma_\mu P_R \otimes \gamma^\mu P_R)\,.
		\end{equation}
		
		\noindent
		Combining all these contributions yields the renormalization constants,
		\begin{eqnarray}
		b_{\underset{prst}{nd}, \underset{prst}{nd}}^1 &=& y_d^2 g_1^2\,,  \quad
		b_{\underset{prst}{nd}, \underset{prst}{nd}}^3 = \frac{4}{3} g_3^2\,, \quad \\
		b_{ \underset{prww}{nu}, \underset{prst}{nd}}^1 &=& -\frac{2}{3}y_d y_u g_1^2\delta_{st}N_c\,, 
		\quad
		b_{ \underset{pr}{\phi  n}, \underset{prst}{nd}}^1 = -\frac{2}{3}y_d y_h g_1^2\delta_{st}\,, 
		\end{eqnarray}
		and subsequently combining these constants with the field renormalizations \eqref{eq:field-constants}, the elements of the ADM are obtained using Eq.~\eqref{eq:adms-master}:
		\begin{eqnarray}
		(\gamma_1)_{\underset{prst}{nd}, \underset{prst}{nd}} &=& -2(-y_d^2+y_d^2 ) g_1^2 =0\,,  \quad
		(\gamma_3)_{\underset{prst}{nd}, \underset{prst}{nd}} = -2(-\frac{4}{3}+\frac{4}{3} )g_3^2 =0\,, \\
		(\gamma_1)_{\underset{prww}{nu}, \underset{prst}{nd}} &=& -2(-\frac{2}{3}y_d y_u \delta_{st} )N_cg_1^2\,,
		\quad
		(\gamma_1)_{ \underset{pr}{ \phi n}, \underset{prst}{nd}} = -2(-\frac{2}{3}y_d y_h \delta_{st} )g_1^2\,,
		\end{eqnarray}
		where $(\gamma_1)_{ij} \equiv -2 g_1^2 \left( \sum_{\psi = \psi_1, ..\psi_4} \frac{1}{2}a^1_\psi \delta_{ij} + b^{1}_{ij}  \right ) $, and similarly for $\gamma_2$ and $\gamma_3$.\\
		
		\begin{boldmath}
			\subsubsection{$\mathcal{O}_{\phi n}$-$\mathcal{O}_{nd}$ mixing}
		\end{boldmath}
		\noindent
		$\mathcal{D}_d$ which involves a penguin insertion 
		of $\mathcal{O}_{nd}$ can be computed using Eq.~(\ref{eq:div:penguin}):
		\begin{eqnarray}
		\mathcal {{D}}_d[B] &=& 
		-\frac{\alpha_1}{4\pi}\frac{1}{\varepsilon}
		(\gamma_\mu P_R \otimes D^\mu P_R) {2 \over 3} y_d y_h N_c\,.
		\end{eqnarray}
		The renormalization constant and anomalous dimension are then
		\begin{equation}
		b^1_{\underset{prst}{nd}, \underset{pr }{\phi n}} = -\frac{2}{3}y_d y_h N_c g_1^2 \delta_{st}\,,	
		\end{equation}
		\begin{equation}
		(\gamma_1)_{\underset{prst}{nd}, \underset{pr}{\phi n}} = \frac{4}{3}y_d y_h N_c g_1^2 \delta_{st}\,.
		\end{equation}
		Note that in this case there are no contributions from wavefunction renormalization.
		\\

		\begin{boldmath}
		\subsubsection{$\mathcal{O}_{nedu}$-$\mathcal{O}_{nedu}$ mixing}
		\end{boldmath}
		\noindent
		For the insertion of $\mathcal{O}_{nedu}=(\bar n_p \gamma_\mu e_r)(\bar d_s \gamma^\mu u_t)$, we have
		\begin{eqnarray}
		\psi_1 &=& n_p\,, \quad \psi_2= e_r\,,  \quad \psi_3 = d_s\,,  \quad \psi_4= u_t\,, \\
		\hat V_1 &\otimes& \hat{V_2} = \delta_{\alpha \beta}\,, 
		\quad \Gamma_1 \otimes \Gamma_2= \gamma_\mu P_R\otimes \gamma^\mu P_R\,.
		\end{eqnarray}
		In this case, $\mathcal{D}_a^{(2)}$, $\mathcal{D}_b^{(2)}$ 
		and $\mathcal{D}_c^{(2)}$ mediated by the gauge boson $B_\mu$, and $\mathcal{D}_a^{(2)}$  
		mediated by $G_\mu$ are generated. We find
		\begin{eqnarray}
		%%%%%%%%
		\mathcal{D}_a^{(2)}[B] &=& \frac{\alpha_1}{4\pi} \frac{1}{4} \frac{1}{\varepsilon} (4y_d y_u) (\gamma_\mu P_R \otimes \gamma^\mu P_R) \,,\\
		%%%%%%%%%
		\mathcal{D}_b^{(2)}[B] &=& -\frac{\alpha_1}{4\pi} \frac{1}{4} \frac{1}{\varepsilon} (16y_e y_u) (\gamma_\mu P_R \otimes \gamma^\mu P_R) \,,\\
		%%%%%%%%%
		\mathcal{D}_c^{(2)}[B] &=& \frac{\alpha_1}{4\pi} \frac{1}{4} \frac{1}{\varepsilon} (4y_d y_e) (\gamma_\mu P_R \otimes \gamma^\mu P_R) \,,\\
		%%%%%%%%%	
		\mathcal{D}_a^{(2)}[G] &=& \frac{\alpha_3}{4 \pi} \frac{1}{4}\frac{1}{\varepsilon} \frac{16}{3}  (\gamma_\mu P_R \otimes \gamma^\mu P_R) \,.
		%%%%%%%%%
		\end{eqnarray}
		Therefore, the divergent parts are 
		\begin{eqnarray}
		b_{\underset{prst}{nedu},\underset{prst}{nedu}}^1 = (y_d y_u - 4 y_e y_u+ y_d y_e)g_1^2\,, 
		\quad b_{\underset{prst}{nedu},\underset{prst}{nedu}}^{3} = \frac{4}{3} \,,
		\end{eqnarray}
		and  using Eqs.~\eqref{eq:field-constants} and \eqref{eq:adms-master}, the elements of the ADM are
		\begin{equation}
		(\gamma_1)_{\underset{prst}{nedu},\underset{prst}{nedu}}
		= -2\left (-\frac{y_d^2}{2}-\frac{y_u^2}{2}-\frac{y_e^2}{2}+ y_d y_u - 4 y_e y_u+ y_d y_e \right ) g_1^2\,,\, 
		(\gamma_3)_{\underset{prst}{nedu},\underset{prst}{nedu}}=-2(-\frac{4}{3}+\frac{4}{3})=0\,. 
		\end{equation}\\
		
		\begin{boldmath}
			\subsubsection{$\mathcal{O}_{\ell n \ell e}$-$\mathcal{O}_{\ell n \ell e}$ mixing}
		\end{boldmath}
		\noindent
		The operator 
		$\mathcal{O}_{\ell n \ell e}=(\bar \ell_p^j n_r)\epsilon_{jk} (\bar \ell_s^k e_t)$ 
		mixes with itself through its insertion into the diagrams 
		$\mathcal{D}_a^{(2)}[B]$, $\mathcal{D}_b^{(2)}[B]$, $\mathcal{D}_c^{(2)}[B]$ and 
		$\mathcal{D}_b^{(2)}[W]$. We have
		\begin{eqnarray}
		\psi_1 &=& \ell_p^j\,, \quad \psi_2= n_r\,,  \quad \psi_3 = \ell_s^k\,,  \quad \psi_4= e_t\,, \\
		\hat V_1 &\otimes & \hat{V_2} = \varepsilon_{jk} \,, \quad 
		\Gamma_1 \otimes \Gamma_2= P_R \otimes  P_R\,.
		\end{eqnarray}
		The contributions to the divergent parts are
		\begin{eqnarray}
		%%%%%%%%
		\mathcal{D}_a^{(2)}[B] &=& \frac{\alpha_1}{4\pi} \frac{1}{4}\frac{1}{\varepsilon} 
		\left ( (P_R) \otimes \gamma_\alpha \gamma_\beta (P_R) \gamma^\beta \gamma^\alpha \right )
		y_\ell y_e \varepsilon_{jk} \,,\\
		%%%%%%%%%
		\mathcal{D}_b^{(2)}[B] &=& -\frac{\alpha_1}{4\pi} \frac{1}{4}\frac{1}{\varepsilon} 
		\left ( \gamma_\alpha \gamma_\beta (P_R)
		\otimes \gamma^\alpha \gamma^\beta (P_R) \right ) y_\ell^2 \varepsilon_{jk} \,,\\
		%%%%%%%%%
		\mathcal{D}_c^{(2)}[B] &=& \frac{\alpha_1}{4\pi} \frac{1}{4}\frac{1}{\varepsilon} 
		\left (\gamma_\alpha \gamma_\beta (P_R)
		\otimes (P_R) \gamma^\beta \gamma^\alpha \right )y_\ell y_e \varepsilon_{jk} \,,\\
		%%%%%%%%%       
		\mathcal{D}_b^{(2)}[W] &=& -\frac{\alpha_2}{4\pi} \frac{1}{4}\frac{1}{\varepsilon}
		\left (\gamma_\alpha \gamma_\beta (P_R)
		\otimes \gamma^\alpha \gamma^\beta (P_R)  \right ){\frac{3}{4}}\varepsilon_{jk} \,.
		%%%%%%%%%
		\end{eqnarray}
		
		After simplification using the Fierz identity,
		\beq
		(\bar{\ell}_p^j \sigma^{\mu\nu} P_R n_r)\epsilon_{jk}(\bar{\ell}^k_s \sigma_{\mu\nu} P_R e_t) = - 4 \mathcal{O}_{\underset{prst}{\ell n \ell e}} + 8 \mathcal{O}_{\underset{srpt}{\ell n \ell e}}\,,
		\eeq
		we find the renormalization constants to be
		\begin{eqnarray}
		b_{\underset{prst}{\ell n \ell e}, \underset{prst}{\ell n \ell e} }^1 &=& (4y_\ell y_e -2 y_\ell^2)g_1^2 \,, \quad  
		b_{ \underset{srpt}{\ell n \ell e}, \underset{prst}{\ell n \ell e}}^1 = (2 y_\ell^2 + 2 y_\ell y_e) g_1^2\,,  \\
		b_{\underset{prst}{\ell n \ell e}, \underset{prst}{\ell n \ell e} }^2 &=& \frac{3}{2}g_2^2 \,, \quad  
		b_{\underset{srpt}{\ell n \ell e}, \underset{prst}{\ell n \ell e} }^2 = - \frac{3}{2} g_2^2.\,  
		\end{eqnarray}
		The elements of the ADM are
		\begin{eqnarray}
		(\gamma_1)_{\underset{prst}{\ell n \ell e}, \underset{prst}{\ell n \ell e}} &=&
		-2 (-3 y_\ell^2 - \frac{y_e^2}{2}  + 4 y_\ell y_e)g_1^2\,,  \quad
		(\gamma_2)_{\underset{prst}{\ell n \ell e}, \underset{prst}{\ell n \ell e}} = -2 (-\frac{3}{4} + \frac{3}{2}) g_2^2\,, \\
		%%%%%%%%%%
		(\gamma_1)_{\underset{srpt}{\ell n \ell e}, \underset{prst}{\ell n \ell e}} &=&
		-2 (2 y_\ell^2 + 2 y_\ell y_e) g_1^2\,, \quad
		(\gamma_2)_{ \underset{srpt}{\ell n \ell e}, \underset{prst}{\ell n \ell e}} = -2\, (-\frac{3}{2}) g_2^2\,.
		\end{eqnarray}

	\end{appendix}
 
%%%%%%%%%%%%%%%%%%%%%%%%%%%%%%%%%%%%%%%%%%%%%%%
%%
\newpage
\bibliographystyle{JHEP}
\bibliography{ref}

\end{document}